\documentclass[aps,showpacs,onecolumn,10pt, notitlepage]{revtex4}
\usepackage{amsfonts}
\usepackage{amsmath}
\usepackage{amssymb}
\usepackage{graphicx}
\usepackage[latin1]{inputenc}
\usepackage[T1]{fontenc}
\usepackage{color}

\begin{document}

\title{Nonlinear wave interaction and spin models in the MHD regime}
\author{G. Brodin, J. Lundin, J. Zamanian and M. Stefan}
\affiliation{Deptartment of Physics, Umea University SE 901 87 Umea, Sweden}

\begin{abstract}
Here we consider the influence on the electron spin in the MHD regime.
Recently developed models which include spin-velocity correlations are taken
as a starting point. A theoretical argument is presented, suggesting that in
the MHD regime a single fluid electron model with spin correlations is
equivalent to a model with spin-up and spin-down electrons constituting
different fluids, but where the spin-velocity correlations are omitted.
Three wave interaction of 2 shear Alfv\'en waves and a compressional Alfv\'en
wave is then taken as a model problem to evaluate the asserted equivalence.
The theoretical argument turns out to be supported, as the predictions of
the two models agree completely. Furthermore, the three wave coupling
coefficients obey the Manley-Rowe relations, which give further support to
the soundness of the models and the validity of the assumptions made in the
derivation. Finally we point out that the proposed two-fluid model can be
incorporated in standard Particle-In-Cell schemes with only minor
modifications.
\end{abstract}

\maketitle

\affiliation{Department of Physics, Umea University, SE--901 87 Umea,
Sweden}

%____________________________________________________________________________

\section{Introduction}

Considerable interest has recently been devoted to the study of quantum
plasmas, see e.g.\ Refs \cite%
{Shukla-Eliasson,Manfredi2005,Padma-Nature,Garcia2005,Misra2006,Haas2008,Cro-2008,haas2005}%
. Much of the research has been motivated by applications to e.g.\ quantum
wells \cite{Manfredi-quantum-well}, plasmonics \cite{Atwater-Plasmonics},
spintronics \cite{Spintronics}, astrophysics \cite{Astro} and ultra-cold
plasmas \cite{Ultracold}. Two of the most basic and much studied quantum
effects are those of the Fermi pressure and the particle dispersive effects
(directly associated with the Bohm De Broglie potential), see e.g.\ Refs.\ \cite{Shukla-Eliasson,Manfredi2005,Padma-Nature,Garcia2005,Misra2006,Haas2008,Cro-2008,haas2005}. Other studies 
\cite%
{Padma-Nature,Asenjo,Vladimirov,Lundin2010,brodin2007-1,brodin2008,brodin2010,zamanian2010,zamanian2010-2}
focus on the electron spin properties that result in the magnetic dipole
force and a magnetization current, in addition to some more complex aspects
of the spin dynamics. Although most quantum effects has a tendency to be
more important in plasmas of high density and low temperature, the regimes
of relevance differ to some extent for the various quantum effects, see
Ref.\ \cite{Lundin2010} for a discussion of this issue. A consequence is
that it is possible to focus on certain of the quantum effects and ignore
the others. In this paper we will make use of this fact and concentrate on
the physics associated with the spin coupling in the Pauli Hamiltonian. Our
starting point is a recenty presented spin-fluid model \cite{zamanian2010-2}%
, derived from kinetic theory \cite{zamanian2010}, which in addition to the
most basic spin precession dynamics includes effects of spin-velocity
correlations. Evaluating this model in the MHD regime, we make a conjecture
based on certain teoretical arguments; That there are two equivalent ways to
model spin-MHD dynamics, either by a one-fluid model including spin-velocity
correlations, or by a two-fluid model without spin-velocity correlations. In
the latter case the spin-up and spin-down states relative to the magnetic
field are regarded as different fluids \cite{brodin2008}. The conjectured
equivalence of these models in the MHD regime is tested by considering a
specific problem of three-wave interaction. For this purpose we calculate
the coupling coefficients between two shear Alfv\'en waves and one
compressional Alfv\'en wave (fast magnetosonic wave) in a magnetized plasma.
The coupling coefficients turns indeed out to be identical in the two cases,
and the coefficients are also seen to obey the Manley-Rowe relations, which
give further support to the soundness of the models used. The applicability
of the two-fluid model \textit{without spin-velocity correlations in the
MHD-regime, }which is strongly supported by our findings, is a very useful
result. The reason is that as this model can be easily adopted into standard
Particle-In-Cell schemes with only small modifications, as will be discussed
in the final section.

\section{Model equations}

Starting from a scalar kinetic equation for a spin-1/2 particle \cite%
{zamanian2010}, spin fluid equations can be derived \cite{zamanian2010-2}.
These are given by the continuity equation 
\begin{equation}
\partial _{t}n^{(s)}+\nabla \cdot (n^{(s)}\mathbf{v}^{(s)})=0
\label{eq:Continuity}
\end{equation}%
and the fluid momentum equation 
\begin{equation}
m^{(s)}\frac{Dv_{i}^{(s)}}{Dt}=q^{(s)}\left( E_{i}+\varepsilon
_{ijk}v_{j}^{(s)}B_{k}\right) +\mu ^{(s)}S_{j}^{(s)}\frac{\partial B_{j}}{%
\partial x_{i}}-\frac{1}{n^{(s)}}\frac{\partial P_{ij}^{(s)}}{\partial x_{j}}%
,  \label{eq:Momentum}
\end{equation}%
where the superscript $s=e,i$ denotes the species (electrons or ions), $%
D/Dt\equiv \partial _{t}+\mathbf{v}^{(s)}\cdot \nabla $, and summation over
repeated indices $i,j,k=x,y,z$ is implied. Here $m^{(s)}$ is the mass, $%
q^{(s)}$ is the charge, $\mu ^{(s)}$ the magnetic dipole moment, $n^{(s)}$
is the number density, and $\mathbf{v}^{(s)}$ is the fluid velocity of
species $s$. Furthermore, $\mathbf{S}$ is the spin vector normalized to
unity, $P_{ij}$ is the pressure tensor, and $\varepsilon_{ijk}$ is the
Levi-Civita symbol. Since the ions normally have a much smaller magnetic
moment than the electrons \cite{proton-note}, the spin contribution due to
the ions can be neglected compared to the electron contribution, i.e. we may
let $\mu ^{(i)}\approx 0$. We have here neglected contributions to the force
from the Fermi pressure and particle dispersive effects (the so called Bohm
de Broglie potential), see \cite{haas2005}. For a discussion of the
parameter regime of importance of various quantum effects, see e.g.\ Refs.\ 
\cite{Quant-classical,Lundin2010}. In this approximation, the pressure moment
satisfies the evolution equation 
\begin{align}
\frac{DP_{ij}^{(s)}}{Dt}=& -P_{ik}^{(s)}\frac{\partial v_{j}^{(s)}}{\partial
x_{k}}-P_{jk}^{(s)}\frac{\partial v_{i}^{(s)}}{\partial x_{k}}-P_{ij}^{(s)}%
\frac{\partial v_{k}^{(s)}}{\partial x_{k}}+\frac{q^{(s)}}{m^{(s)}}%
\varepsilon _{imn}P_{jm}^{(s)}B_{n}+\frac{q^{(s)}}{m^{(s)}}\varepsilon
_{jmn}P_{im}^{(s)}B_{n}  \notag \\
& +\frac{\mu ^{(s)}}{m^{(s)}}\Sigma _{ik}\frac{\partial B_{k}}{\partial x_{j}%
}+\frac{\mu ^{(s)}}{m^{(s)}}\Sigma _{jk}\frac{\partial B_{k}}{\partial x_{i}}%
,  \label{Pressure}
\end{align}%
where again the last two terms can be dropped for the ion species.

Furthermore, to describe the spin dynamics we need the electron spin
evolution equation, which is given by 
\begin{equation}
\frac{DS_{i}^{(e)}}{Dt}=\frac{2\mu ^{(e)}}{\hbar }\varepsilon
_{ijk}S_{j}^{(e)}B_{k}-\frac{1}{m^{(e)}n^{(e)}}\frac{\partial \Sigma
_{ij}^{(e)}}{\partial x_{j}}  \label{eq:Spin}
\end{equation}%
where $\Sigma _{ij}^{(e)}$ is the spin-velocity correlation tensor. Finally
the evolution of the spin-velocity moment is described by 
\begin{align}
\frac{D\Sigma _{ij}^{(e)}}{Dt}=& -\Sigma _{ij}^{(e)}\frac{\partial
v_{k}^{(e)}}{\partial x_{k}}-\Sigma _{ik}^{(e)}\frac{\partial v_{j}^{(e)}}{%
\partial x_{k}}-P_{jk}^{(e)}\frac{\partial S_{i}^{(e)}}{\partial x_{k}} 
\notag \\
& +\frac{q^{(e)}}{m^{(e)}}\varepsilon _{jkl}\Sigma _{ik}^{(e)}B_{l}+\frac{%
2\mu ^{(e)}}{\hbar }\varepsilon _{ikl}\Sigma _{kj}^{(e)}B_{l}+\mu
^{(e)}n^{(e)}\frac{\partial B_{i}}{\partial x_{j}}-\mu
^{(e)}n^{(e)}S_{i}^{(e)}S_{k}^{(e)}\frac{\partial B_{k}}{\partial x_{j}}
\label{eq:SigmaA}
\end{align}%
In Eq.\ (\ref{Pressure}) we have neglected the heat flux tensor $Q_{ijk}$ to
obtain a closed set of equations. Similarly we have neglected the higher
order tensor $\Lambda _{ijk}$ in the evolution equation for the
spin-velocity tensor Eq.\ (\ref{eq:SigmaA}). The validity of the truncation
has been investigated in Refs.\ \cite{zamanian2010,stefan2011-1}, and the
truncation seem to be an accurate approximation in the low-temperature
limit. The equations above together with Maxwell's equations constitute a
closed system, where a magnetization current density $\mathbf{j}_M = \nabla
\times \mathbf{M}$, due to the spin, should be added to the free current
density, and where naturally all species contribute in the latter term. The
set of equations (\ref{eq:Continuity})-(\ref{eq:SigmaA}) has been studied by
Refs.\ \cite{stefan2011-1,zamanian2010-2}, but without inclusion of the ion
dynamics. The aim of the current paper is to apply the above set of
equations to the MHD regime where the ion dynamics is essential, at the same
time as making a careful evaluation of the electron spin magnetization.

\subsection{MHD-limit}

As concluded in the previous section, we will primarily consider wave
dynamics in the MHD regime where the frequencies are smaller than the
ion-cyclotron frequency and the wavelengths are longer than the Larmor
radius. Furthermore, we will consider the low-temperature (i.e. low-beta)
limit, where the pressure terms are dropped. Under these assumptions, the
system will be described by the magnetohydrodynamic equation \cite%
{brodin2007-1} 
\begin{equation}
\rho \left( \frac{\partial }{\partial t}+\mathbf{u}\cdot \nabla \right) 
\mathbf{u}=-\nabla \left( \frac{B^{2}}{2\mu _{0}}-\mathbf{M}\cdot \mathbf{B}%
\right) +(\mathbf{B}\cdot \nabla )\mathbf{M}-\nabla P_{e}  \label{eq:MHD}
\end{equation}%
together with the equations 
\begin{equation}
\partial _{t}\mathbf{B}=\nabla \times \left( \mathbf{u}\times \mathbf{B}%
\right) .  \label{eq:B-MHD}
\end{equation}%
and 
\begin{equation}
\frac{\partial \rho }{\partial t}+\nabla \cdot \left( \rho \mathbf{u}\right)
= 0  \label{eq:Cont-MHD}
\end{equation}%
Here we have neglected the electron contribution to the fluid mass density
by setting $\rho \approx m^{(i)}n^{(i)}$, and the fluid velocity can be
written as $\mathbf{u}=(n^{(e)}m^{(e)}\mathbf{v}^{(e)}+n^{(i)}m^{(i)}\mathbf{%
v}^{(i)})/\rho \approx \mathbf{v}^{(i)}$. We have neglected the magnetic
moment of the ions such that the magnetization $\mathbf{M}=\mu ^{(e)}n^{(e)}%
\mathbf{S}^{(e)}$ is purely due to the electron spin. The derivation of Eq.\
(\ref{eq:MHD}) made in Ref.\ \cite{brodin2007-1} was done starting from a
somewhat less elaborate set of equations, not including the spin-velocity
correlations. However, the derivation does not involve the Eqs.\ (\ref%
{eq:Spin})-(\ref{eq:SigmaA}) describing the spin dynamics, and hence we may
adopt this result within the current model.

Without the magnetization $\mathbf{M}$ we obtain standard ideal MHD
equations, and thus (\ref{eq:MHD}), (\ref{eq:B-MHD}) and (\ref{eq:Cont-MHD})
constitute a closed system. The magnetization can be easily included with
the spin determined by Eqs.\ (\ref{eq:Spin}) together with (\ref{eq:SigmaA}%
), where the terms containing derivatives of the velocity turns out to be
negligible. The aim is to solve for the magnetization in terms of the
magnetic field, in which case Eqs.\ (\ref{eq:MHD}), (\ref{eq:B-MHD}) and (%
\ref{eq:Cont-MHD}) are sufficient to produce a closed spin-MHD theory. This
can be achieved in two different ways; either by considering the electrons
in spin up and spin down states relative the magnetic field as two separate
fluids, or by treating them as a single fluid with a macroscopic spin that
is proportional to the difference in population density of the two spin
states. We will now go on to discuss this in further detail.

%____________________________________________________________________________

\subsection{One-fluid vs two-fluid}

In this sub-section we will discuss how to determine the magnetization, in
order to use Eqs.\ (\ref{eq:MHD}), (\ref{eq:B-MHD}) and (\ref{eq:Cont-MHD}).
To find the magnetization we first need to solve (\ref{eq:Spin}) to
determine the spin. The first term of the right hand side of Eq.\ (\ref%
{eq:Spin}) is the basic spin precession. If the spin-velocity correlations
in (\ref{eq:Spin}) can be neglected, the solutions for $\mathbf{S}$ are
particularly simple in the MHD regime. This is because the left hand side
term of Eq.\ (\ref{eq:Spin}) is smaller than the spin \ precession term by a
factor of the order $O(\omega ^{(\mathrm{ch})}/\omega _{cg}^{(\mathrm{ch})})$%
, where $\omega ^{(\mathrm{ch})}\sim \partial _{t}$ is a characteristic
frequency scale of the problem, and $\omega _{cg}^{(\mathrm{ch})}\sim 2\mu
_{e}B/\hbar $ is the characteristic spin precession frequency (which is
close to the characteristic cyclotron frequency $\omega _{c}^{(\mathrm{ch}%
)}\sim qB/m$). Assuming that spin-velocity correlations can be omitted, we
note that spin evolution equation in the MHD regime reduces to 
\begin{equation}
\varepsilon _{ijk}S_{j}^{(e)}B_{k}=0.  \label{Eq-spin-precession}
\end{equation}%
This has two solutions, where $\mathbf{S}$ is either parallel or
antiparallel to $\mathbf{B}\,$, that is $S_{i}=\pm b_{i}$, where $%
b_{i}=B_{i}/B$ is a unit vector in the direction of $\mathbf{B}$. A
comparatively general way to deal with spins obeying $S_{i}=\pm b_{i}$ is to
consider a two-fluid model of electrons, where for one of the species the
electron spin state is parallel to $\mathbf{B}$, and for the other species
antiparallel. Eq.\ (\ref{Eq-spin-precession}) then implies that these spin
states are conserved. However, as seen from the above discussion this is
only an adequate approximation if the spin-velocity correlations give a
small contribution in Eq.\ (\ref{eq:Spin}). Thus our next step is to outline
the solutions of Eq.\ (\ref{eq:SigmaA}) using MHD approximations, in order
to determine the contribution from $\Sigma _{ij}^{(e)}$ in (\ref{eq:Spin}).
Firstly we note that the three first terms in (\ref{eq:SigmaA}) are at most
of order $\omega ^{(\mathrm{ch})}\Sigma _{ij}$, whereas the fifth and sixth
are of order $\omega _{c}^{(\mathrm{ch})}\Sigma _{ij}$ $\sim \omega _{cg}^{(%
\mathrm{ch})}\Sigma _{ij}$. Thus neglecting the three first terms we can
formally write Eq.\ (\ref{eq:SigmaA}) on the form $\overleftrightarrow{O}%
\cdot \overrightarrow{\Sigma }=\overrightarrow{\sigma }$, where $%
\overleftrightarrow{O}$ is a $9\times 9$-matrix where all coefficients are $%
\pm \omega _{c\alpha }$ or $\pm \omega _{cg\alpha }$. Here $\overrightarrow{%
\Sigma }$ is a $9$-component vector containing all elements of $\Sigma _{ij}$%
, and $\overrightarrow{\sigma }$ is a $9$-component vector containing the
source terms, i.e.\ $P_{jk}(\partial S_{i}/\partial x_{k})$, $\mu n(\partial
B_{i}/\partial x_{j})$ and $\mu nS_{i}S_{k}(\partial B_{k}/\partial x_{j})$, 
$\omega _{c\alpha }=qB_{\alpha }/m$ and $\omega _{cg\alpha }=2\mu
_{e}B_{\alpha }/\hbar $ with $\alpha =x,y,z$. Note that the full field
strength is used and not the linearized field when defining $\omega
_{c\alpha }$ and $\omega _{cg\alpha }$. Since $\overleftrightarrow{O}$
contains no operators, we can do a simple matrix inversion to find $%
\overrightarrow{\Sigma }=\overleftrightarrow{O}^{-1}\overrightarrow{\sigma }$%
. This turns out to be sufficient to determine all components of $\Sigma
_{ij} $, except for a component directed as $\mathbf{b}\otimes \mathbf{b}$.
Thus this approximation scheme allows us to compute $\Sigma _{ij}$ apart
from a contribution that can be expressed as $\Sigma _{ij}=\Phi b_{i}b_{j}$,
where $\Phi $ is a scalar field. Using the solution $\overrightarrow{\Sigma }%
=\overleftrightarrow{O}^{-1}\overrightarrow{\sigma }$ we can easily check
that the determined components of $\Sigma _{ij}$ are of order $\Sigma _{ij}$ 
$\sim \mu _{B}nkB/\omega _{c}^{(\mathrm{ch})}$. This means that the
contributions from $\Sigma _{ij}$ are sufficiently small to be neglected in (%
\ref{eq:Spin}). However, in general we must also account for the
contribution $\Sigma _{ij}=\Phi b_{i}b_{j}$ whose magnitude is unknown.
Since the scalar field $\Phi $ cannot be determined if the three first terms
of (\ref{eq:SigmaA}) are omitted, we must extend our model to the solve the
full case of Eq.\ (\ref{eq:SigmaA}). We will do so within a one-fluid model
in the section III A, and it turns out that $\Phi $ becomes sufficiently
large for this component of $\Sigma _{ij}$ to signficantly influence the
solutions to (\ref{eq:Spin}), also within the MHD-regime. However, it also
turns out that in order to get a large value of $\Phi $, we must have a spin
vector that is different from $\pm \mathbf{b}$. This is the normal case in a
one-fluid theory, where the macroscopic spin results from averaging over all
spin states. However, within a two-fluid MHD model (treating spin-up and
down states as different species) without spin-velocity correlations Eq.\ (%
\ref{Eq-spin-precession}) can be applied leading to $S_{i}=\pm b_{i}$, and
the situation would then again be modified. Indeed, contracting Eq.\ (\ref%
{eq:SigmaA}) with $b_{i}b_{j}$ to compute the source terms for $\Phi $, we
find that all the source terms for $\Sigma _{ij}$ vanishes if $S_{i}=\pm
b_{i}$, as the fourth term in Eq.\ (\ref{eq:SigmaA}) becomes $%
b_{i}b_{j}P_{jk}(\partial S_{i}/\partial x_{j})$, which is zero as $%
b_{i}(\partial S_{i}/\partial x_{j})=\pm (1/2)\partial (b_{i}b_{i})/\partial
x_{j}$, whereas terms 7 and 8$\,$together become 
\begin{equation*}
b_{i}b_{j}\frac{\partial B_{i}}{\partial x_{j}}-b_{i}b_{j}S_{i}S_{k}\frac{%
\partial B_{k}}{\partial x_{j}}=0
\end{equation*}%
where $S_{i}=\pm b_{i}$ was used in the last step. This result provides the
theoretical basis for adopting a two-fluid model of electrons in the
MHD-regime and omitting spin-velocity correlations in (\ref{eq:Spin}),
leading to $S_{i}=\pm b_{i}$. The division into two fluids leaves the rest
of the basic equations structurally unaffected, but we now obatin two
contributions such that the magnetization is calculated as $\mathbf{M}=\mu
n_{\uparrow }\mathbf{s}_{\uparrow }+\mu n_{\downarrow }\mathbf{s}%
_{\downarrow }$ due to the difference in density perturbations of the two
spin states. We will consider this in more detail within perturbation theory
in our model problem below. The conclusions of this section is then
confirmed, since the one-fluid models that keeps the spin-velocity
correlations in Eq.\ (\ref{eq:Spin}) give indeed an identical expression for
the magnetization as the two-fluid model with up and down spins $S_{i}=\pm
b_{i}$. The allowance for independent density variations of the two species
in the latter model provides the physical mechanism that reproduces the
effects of spin-velocity correlations in the one-fluid model. It should
however be stressed that this conclusion is limited to the MHD regime. 
%____________________________________________________________________________

\section{Three wave interaction - a model problem}

We will now consider a model problem with the purpose of testing our
conclusions about the similarities between the one-fluid and two-fluid
models outlined in the previous section. Specifically, we consider three
wave interaction between two shear Alfv\'en waves ($A,A^{\prime }$) and one
compressional Alfv\'en wave ($MS$); $MS\rightarrow A+A^{\prime }$. Using
three-wave interaction as a model problem has the advantage that an
unphysical assumption (or an incorrect calculation) is likely to result in a
broken Manley-Rowe symmetry \cite{Manley-Rowe}, in which case one gets a
clear indication that something needs to be revised.

The waves are assumed to be small perturbations on a homogeneous background,
and we write $\mathbf{B}=B_{0}\hat{\mathbf{z}}+\mathbf{B}_{1}$, $\rho =\rho
_{0}+\rho _{1}$, etc., but omit the index 1 on variables whose backgound
values are zero. Furthermore, we omit index 1 whenever the cartesian
components are specified for notational convenience, i.e. we write $\mathbf{B%
}_{1}=B_{x}\hat{\mathbf{x}}+B_{y}\hat{\mathbf{y}}+B_{z}\hat{\mathbf{z}}$.
Moreover, we assume that there is no drift so that $\mathbf{u}_{0}=0$ and
also that there is no spin-velocity correlation in the background
distribution, i.e. $\Sigma _{0}=0$. For simplicity we further assume that
the temperature is sufficiently low such that the equilibrium pressure can
be neglected, $P_{0}=0 $ (i.e. that we have a low-beta plasma with the
ion-acoustic velocity much smaller than the Alfv\'en velocity). Furthermore,
assuming that $\mu _{B}B_{0}/(k_{B}T)\ll 1$ we can make the approximation
that the equilibrium spin up and spin down populations are equal so that $%
n_{0\uparrow }=n_{0\downarrow }$ in the two-fluid model which implies that
the total zeroth order magnetization vanishes \cite{Magnetization-note}. For
consistency between the one-fluid and two-fluid models we should
consequently pick $\mathbf{M}_{0}=0$ also in the latter case. The difference
in the model equations between the one-fluid and two-fluid approach is then
primarily that in the two-fluid model we have $\mathbf{S}_{0}=\pm \hat{%
\mathbf{z}}$ for the two spin states (in which case we obtain a finite zero
order magnetization if only one of the electron fluids are counted) whereas
in the one-fluid model $\mathbf{M}_{0}=0$ and $\mathbf{S}_{0}=0$.

Next we make a harmonic decomposition $\partial _{t}\rightarrow -i\omega $
and $\partial _{\mathbf{x}}\rightarrow i\mathbf{k}$ for each wave, where the
frequencies and wave vectors satisfy the conditions 
\begin{align}
\omega ^{MS}& =\omega ^{A}+\omega ^{A^{\prime }}  \label{matching1} \\
\mathbf{k}^{MS}& =\mathbf{k}^{A}+\mathbf{k}^{A^{\prime }}.  \label{matching2}
\end{align}%
with the index $MS$ denoting the compressional Alfv\'en (or fast magnetosonic)
wave, and the index $A$ and $A^{\prime }$ denoting the shear Alfv\'en waves.
The coordinate system is defined so that the $z$-direction points in the
direction of the unperturbed magnetic field, $\mathbf{B}=B_{0}\hat{\mathbf{z}%
}$, and for simplicty we assume all wave vectors to lie in the $xz$-plane.

Throughout the calculation we will use $\omega /\omega _{c}$ and $%
kC_{A}/\omega _{c}$ as small expansion parameters (where $\omega
_{c}=qB_{0}/m$ is the cyclotron frequency), in accordance with standard MHD
theory. Here $C_{A}=(B_{0}^{2}/\mu _{0}m_{i}n_{0})^{1/2} =
(B_0^2/\mu_0\rho_0)^{1/2}$ is the Alfv\'en velocity, and $\omega $ and $k$
represents any of the wave frequencies or wave vector components. We also
note that $\omega _{cg}\simeq \omega _{c}$, where $\omega _{cg}=$ $2\mu
_{e}B_{0}/\hbar $ is the spin precession frequency. Furthermore, $\omega
_{cg}-\omega _{c}$ is of the same order as the ion-cyclotron frequency, and
is therefore much larger than wave frequencies within the MHD regime \cite%
{Frequency-note}. We will therefore drop terms proportional to $(\omega
_{cg}-\omega _{c})^{-1}$ compared to $\omega ^{-1}$ in our final results.

It should be pointed out that unlike the classical case, the pressure tensor 
$P_{ij}$ does not necessarily vanish in the limit of zero temperature.
However, we note that we need not be concerned about the contribution from
the pressure term in this particular case. This is because $P_{ij}$ vanishes
linearly in the MHD limit and thereby enters as a cubic nonlinearity (which
does not affect the three-wave interaction) in the evolution equation for $%
\Sigma _{ij}$. The pressure tensor, however, also gives a contribution in
the MHD equation \eqref{eq:MHD}, but it turns out that this is a nonlinear
contribution proportional to $(\omega _{cg}-\omega _{c})^{-1}$ which is
small compared to leading terms proportional to $\omega ^{-1}$. We may
therefore neglect the contribution from $P_{ij}$ altogether.

%%%%%%%%%%%%%%%%%%%%%%%%%%%%%%%%%%%%%%%%%%%%%

\subsection{One-fluid calculation}

We start by considering the one fluid model for which, as mentioned above,
the unperturbed spin-density is zero, $\mathbf{S}_{0}=0$. Our first aim is
to find the linear dispersion relation as well as the linear eigenvectors
(polarizations) of the shear Alfv\'en wave and the compressional Alfv\'en wave.
We note that in the MHD equation \eqref{eq:MHD} we need an expression for
the magnetization. We therefore begin by solving the spin-velocity evolution
equation to find the $\Sigma $-tensor. Linearly, this is straightforward and
we find 
\begin{equation}
\Sigma _{ij}=in_{0}\mu \left( 
\begin{array}{ccc}
-\frac{k_{x}B_{y}\omega _{cg}}{\omega _{c}^{2}-\omega _{cg}^{2}} & -\frac{%
k_{x}B_{x}\omega _{c}}{\left( \omega _{c}^{2}-\omega _{cg}^{2}\right) } & 
\frac{k_{z}B_{y}}{\omega _{cg}} \\ 
\frac{k_{x}B_{x}\omega _{cg}}{\left( \omega _{c}^{2}-\omega _{cg}^{2}\right) 
} & -\frac{k_{x}B_{y}\omega _{c}}{\omega _{c}^{2}-\omega _{cg}^{2}} & -\frac{%
k_{z}B_{x}}{\omega _{cg}} \\ 
0 & -\frac{k_{x}B_{z}}{\omega _{c}} & i\frac{k_{z}B_{z}}{\omega }%
\end{array}%
\right) .  \label{eq:linear-sigma-tensor}
\end{equation}%
As can be seen, the components in (\ref{eq:linear-sigma-tensor}) have
different magnitudes, but we keep all of them at this stage in the
calculation. Next we use the linear $\Sigma $-tensor in the spin evolution
equation \eqref{eq:Spin} to find an expression for the linear spin $\mathbf{S%
}$ and thereby the linear magnetization $\mathbf{M}=\mu n_{0}\mathbf{S}$. 
\begin{equation}
\mathbf{M}=\frac{n_{0}\mu ^{2}}{m}\left( 
\begin{array}{c}
\frac{k_{x}^{2}B_{x}}{(\omega _{c}^{2}-\omega _{cg}^{2})} \\ 
\frac{k_{x}^{2}B_{y}}{(\omega _{c}^{2}-\omega _{cg}^{2})} \\ 
-\frac{k_{z}^{2}B_{z}}{\omega ^{2}}%
\end{array}%
\right)  \label{linear-magn}
\end{equation}%
Here we have dropped contributions to components of $\mathbf{M}$ that are
smaller by factors $\left\vert \omega _{c}^{2}-\omega _{cg}^{2}\right\vert
/\omega _{cg}^{2}$ and/or $\omega /\omega _{cg}$. \ Substituting (\ref%
{linear-magn}) into (\ref{eq:MHD}), the linear dispersion relations are
obtained from (\ref{eq:MHD}), (\ref{eq:B-MHD}) and (\ref{eq:Cont-MHD}).
Similarly to the classical ideal MHD case, the modes decouple into the shear
Alfv\'en wave described by%
\begin{equation}
D_{A}(\omega ,\mathbf{k})\equiv \omega ^{2}-k_{z}^{2}C_{A}^{2}\left( 1-\frac{%
n_{0}\mu _{0}\mu ^{2}}{m}\frac{k_{x}^{2}}{(\omega _{c}^{2}-\omega _{cg}^{2})}%
\right) =0  \label{Shear-DR}
\end{equation}%
and the compressional Alfv\'en (fast magnetosonic) wave with the dispersion
relation

\begin{equation}
D_{MS}(\omega ,\mathbf{k})\equiv \omega ^{2}-k_{x}^{2}C_{A}^{2}\left( 1+%
\frac{n_{0}\mu _{0}\mu ^{2}}{m}\frac{k_{z}^{2}}{\omega ^{2}}\right)
-k_{z}^{2}C_{A}^{2}\left( 1-\frac{n_{0}\mu _{0}\mu ^{2}}{m}\frac{k_{x}^{2}}{%
(\omega _{c}^{2}-\omega _{cg}^{2})}\right) =0 .  \label{Compressional-DR}
\end{equation}%
Note that the last terms in (\ref{Shear-DR}) and (\ref{Compressional-DR})
are smaller than the first spin-modification in (\ref{Compressional-DR}) as $%
\omega ^{2}\ll \left\vert \omega _{c}^{2}-\omega _{cg}^{2}\right\vert $.
Furthermore, the linear eigenvector components for the shear Alfv\'en wave are 
$u_{x}^{A}=u_{z}^{A}=0$, $B_{x}^{A}=B_{z}^{A}=0$, $\rho _{1}^{A}=0$, and 
\begin{equation}
B_{y}^{A}=-\frac{k_{z}^{A}B_{0}}{\omega ^{A}}u_{y}^{A}  \label{linear-1}
\end{equation}%
For the compressional Alfv\'en wave (index $MS$) we instead obtain $%
u_{y}^{MS}=u_{z}^{MS}=0$, $B_{y}^{MS}=0$, and 
\begin{eqnarray}
B_{x}^{MS} &=&-\frac{k_{z}^{MS}B_{0}}{\omega ^{MS}}u_{x}^{MS}
\label{linear-ms-1} \\
B_{z}^{MS} &=&\frac{k_{\bot }^{MS}B_{0}}{\omega ^{MS}}u_{x}^{MS}
\label{linear-ms-2} \\
\rho _{1}^{MS} &=&\rho _{0}\frac{k_{\bot }^{MS}}{\omega ^{MS}}u_{x}^{MS}
\label{linear-ms-3}
\end{eqnarray}%
Next we aim to calculate the three wave coupling coefficients due to the
quadratic nonlinearities. We have calculated the nonlinear contribution to
the coupling coefficients including all terms proportional to $(\omega
_{cg}-\omega _{c})^{-1}$. However, our results show that these terms only
give rise to small corrections to the leading terms proportional to $\omega
^{-1}$. Since the full analysis is rather tedious we will therefore only
write out the leading terms in the NL contribution to the $\Sigma $-tensor
as well as to the magnetization $\mathbf{M}$. Under the given
approximations, keeping the resonant terms, we find the components with a
nonzero nonlinear contribution to the $\Sigma $-tensor to be

\begin{equation}
\Sigma _{yz}^{A^{\prime }}=-\frac{k_{z}B_{x}^{A^{\prime }}}{\omega _{cg}}+%
\frac{2i\mu }{\hbar }\frac{k_{z}^{MS}B_{z}^{MS}B_{y}^{A^{\ast }}}{\omega
_{cg}\omega ^{MS}}  \label{Sigma-NLA1}
\end{equation}%
and%
\begin{equation}
\Sigma _{zy}^{A^{\prime }}=-\frac{k_{x}B_{z}}{\omega _{c}}+\frac{iq}{m}\frac{%
k_{z}^{MS}B_{z}^{MS}B_{y}^{A^{\ast }}}{\omega _{c}\omega ^{MS}}
\label{Sigma-NLA2}
\end{equation}%
for the Alfv\'en wave. Here $*$ denotes complex conjugation. For the
magnetosonic wave the component with a nonzero nonlinear contribution is%
\begin{equation}
\Sigma _{33}^{MS}=i\frac{k_{z}B_{z}^{MS}}{\omega }+\frac{2i\mu }{\hbar }%
\frac{1}{\omega _{cg}\omega ^{MS}}\left( k_{z}^{A}+k_{z}^{A^{\prime
}}\right) B_{y}^{A^{\prime }}B_{y}^{A}  \label{Sigma-NL-MS}
\end{equation}%
Solving the spin evolution equation with the sources from $\Sigma $ given
above, we find the a nonlinear contribution to the $z$-component of the
magnetization 
\begin{equation}
M_{z}^{MS}=-\frac{n_{0}\mu ^{2}}{m}\left( \frac{k_{z}^{2}B_{z}^{MS}}{\omega
^{2}}+\frac{2\mu }{\hbar }\frac{k_{z}^{MS}\left( k_{z}^{A}+k_{z}^{A^{\prime
}}\right) }{\omega _{cg}\omega ^{2(MS)}}B_{y}^{A}B_{y}^{A^{\prime }}\right)
\label{NL-magn1}
\end{equation}%
for the MS wave, and a nonlinear contribution to the $y$-component of the
magnetization 
\begin{equation}
M_{y}^{A^{\prime }}=\frac{n_{0}\mu ^{2}}{m}\left( \frac{k_{x}^{2}B_{y}}{%
(\omega _{c}^{2}-\omega _{cg}^{2})}-\frac{2\mu }{\hbar }\frac{k_{z}^{2(MS)}}{%
\omega _{cg}\omega ^{2(MS)}}B_{z}^{MS}B_{y}^{A\ast }\right)  \label{NL-magn2}
\end{equation}%
for the Alfv\'en wave. Now that we have expressed the magnetization in terms
of the magnetic field, correct to second order in the amplitude, we may
substitute these results into (\ref{eq:MHD}), and perform the rest of the
calculations using (\ref{eq:MHD}), (\ref{eq:B-MHD}) and (\ref{eq:Cont-MHD})
as in standard MHD theory. Accounting for time dependent amplitudes with the
substitions $D_{A}(\omega ,\mathbf{k})\rightarrow \lbrack \partial
D_{A}/\partial \omega ]i\partial /\partial t$ and $D_{MS}(\omega ,\mathbf{k}%
)\rightarrow \lbrack \partial D_{MS}/\partial \omega ]i\partial /\partial t$%
, doing successive elimination keeping the velocity variables as the wave
amplitudes, we find the following coupled equations for the different wave
modes 
\begin{equation}
\frac{\partial u_{y}^{A^{\prime }}}{\partial t}=-i\frac{\omega ^{2(A^{\prime
})}}{\partial D_{A^{\prime }}/\partial \omega }Cu_{y}^{A^{\ast }}u_{x}^{MS}
\label{Wave-NL-1}
\end{equation}%
and 
\begin{equation}
\frac{\partial u_{x}^{MS}}{\partial t}=-i\frac{\omega ^{2(MS)}}{\partial
D_{MS}/\partial \omega }Cu_{y}^{A}u_{y}^{A^{\prime }}  \label{Wave-NL-2}
\end{equation}%
with the coupling coefficient 
\begin{equation}
C=\frac{k_{x}^{MS}}{\omega ^{MS}}\left( 1+\frac{n_{0}\mu _{0}\mu ^{2}}{m}%
\frac{k_{z}^{2(MS)}}{\omega ^{2MS}}\right) .  \label{Coeff}
\end{equation}%
Due to the symmetry between the two shear Alfv\'en waves, the equation for $%
\partial u_{y}^{A}/\partial t$ is obtained by exchanging $A$ and $A^{\prime
} $ in Eq.\ (\ref{Wave-NL-1}). The appearance of the common factor $C$ in
the three coupled equations is a reflection of the Manley-Rowe symmetry \cite%
{Manley-Rowe}. The first term of $C$ is a purely classical contribution,
that agrees with e.g.\ Refs.\ \cite{Chin1972,Classical-C} in the cold limit.
For the spin contribution in Eq.\ (\ref{Coeff}) \ to be important as
compared to the classical one, a rather dense plasma is required. By
contrast, other MHD phenomena exists that require less extreme parameters
for the electron spin to be important. Nevertheless, as will be discussed in
the final section, the results derived here have a number of interesting
theroretical consequences. It should be stressed that the contribution to
the magnetization in this one-fluid model stems from the $\Sigma $-tensor.
This is in contrast to the two-fluid model as we will see below.

\subsection{Two-fluid calculation}

We now consider the problem of three wave coupling using the two-fluid
model. The spin is then determined from (\ref{eq:Spin}) with the
contribution from $\Sigma $ omitted, as described in section II, but we now
have two species of electrons, which have the unperturbed spin $\mathbf{S}%
_{0\uparrow }=\hat{\mathbf{z}}$ and $\mathbf{S}_{0\downarrow }=-\hat{\mathbf{%
z}}$, respectively. The total magnetization is then written as 
\begin{equation}
\mathbf{M}=\mu \left( n^{(\uparrow )}\mathbf{S}^{(\uparrow)}+n^{(\downarrow)
}\mathbf{S}^{(\downarrow) }\right)  \label{eq:Magn-2f}
\end{equation}%
which gives $\mathbf{M}_{0}=0$ in agreement with the previous section,
provided we let $n_{0\uparrow }=n_{0\downarrow }=n_{0}/2$ which will be used
henceforth. Next we find the linear spin-vector to be 
\begin{equation}
\mathbf{S}_{1}=\left( 
\begin{array}{c}
\frac{2\mu }{\hbar \omega _{cg}}S_{0}B_{x}+\frac{\mu }{m}\frac{k_{x}^{2}B_{x}%
}{\omega _{c}^{2}-\omega _{cg}^{2}} \\ 
\frac{2\mu }{\hbar \omega _{cg}}S_{0}B_{y}+\frac{\mu }{m}\frac{k_{x}^{2}B_{y}%
}{\omega _{c}^{2}-\omega _{cg}^{2}} \\ 
0%
\end{array}%
\right) .  \label{spin-two-fluid}
\end{equation}%
Note here that although the terms $\propto S_0$ in (\ref{spin-two-fluid})
are larger than the terms $\propto (\omega _{c}^{2}-\omega _{cg}^{2})^{-1}$,
the former has opposite signs for the up- and down species, and hence give
no contribution to the linear magnetization. It turns out that the terms in (%
\ref{spin-two-fluid}) $\propto (\omega _{c}^{2}-\omega _{cg}^{2})^{-1}$ are
needed to get agreement with the linear magnetization obtained with the
one-fluid model \eqref{linear-magn}. However, it can be noted that these
terms have been dropped in the expression for the coupling coefficient %
\eqref{Coeff} where, in the end, only the leading term is kept. Next we need
to find an expression for the fluid densities of the electron spin fluids. From
the continuity equation \eqref{eq:Continuity} we have 
\begin{align}
n^{(s)} = n_0^{(s)} + n_0^{(s)}\frac{\mathbf{k}\cdot\mathbf{v}^{(s)}}{\omega}
+ n^{(s)\text{NL}}
\end{align}
where $n^{(s)\text{NL}}$ is a non-linear contribution that can be shown not
to contribute to the magnetization after summation of the spin states $%
s=\uparrow,\downarrow$. An expression for the electron velocities is
obtained by solving the fluid momentum equation \eqref{eq:Momentum} together
with $-\partial _{t}\mathbf{B}=\nabla \times \mathbf{E}$. To close this
system we may in general need to use a full fluid description. However,
within the MHD regime and for this specific problem, it suffices to
determine the magnetization. For our case with $n_{0\uparrow
}=n_{0\downarrow }=n_{0}/2$ several terms vanish in Eq.\ (\ref{eq:Magn-2f})
after the summation over up and down species. In MHD we make the
approximation that we may write the electric field as $\mathbf{E}=-\mathbf{v}%
^{(i)}\times \mathbf{B\simeq }-\mathbf{u}\times \mathbf{B}$. This allows us
to again make use of Eqs.\ (\ref{linear-1})-(\ref{linear-ms-3}). Under these
assumptions we note that $E_{z}$ vanishes linearly, and also that $E_{z}^{NL}
$ is only proportional to quadratic combinations of $B$-field components and
will therefore not contribute to the magnetization. Thus $E_{z}$ may
therefore be set to zero from now on. Under these assumptions, it is easy to
show that the fluid velocities may be written as 
\begin{subequations}
\label{eq:v}
\begin{align}
v_{x}& =\frac{q}{m}\frac{\omega }{\omega _{c}}\frac{B_{z}}{k_{x}}+\frac{q}{m}%
\frac{1}{\omega _{c}}\left( -v_{x}B_{z}+v_{z}B_{x}\right) \\
v_{y}& =-\frac{q}{m}\frac{\omega }{\omega _{c}}\frac{B_{y}}{k_{z}}+i\frac{%
\mu }{m\omega _{c}}k_{x}B_{z}S_{0}-\frac{q}{m}\frac{1}{\omega _{c}}\left(
v_{y}B_{z}-v_{z}B_{y}\right) -i\frac{\mu }{m\omega _{c}}k_{x}\left(
B_{x}S_{x}-B_{y}S_{y}\right) \\
v_{z}& =-\frac{\mu }{m\omega }k_{z}B_{z}S_{0}+i\frac{q}{m}\frac{1}{\omega }%
\left( v_{x}B_{y}-v_{y}B_{x}\right) -\frac{\mu }{m\omega }k_{z}\left(
B_{x}S_{x}-B_{y}S_{y}\right)
\end{align}%
\end{subequations}
From Eqs.\ \eqref{eq:Magn-2f}--\eqref{eq:v} we find the linear
magnetization, and it agrees identically with the expression obtained from
the one-fluid model. Furthermore, an extended analysis gives agreement for
the NL terms of the magnetization as well. \ Consequently, the coupling
coefficients remain the same regardless if the one-fluid or two-fluid model
are used to determine the magnetization. This corroborates the usefulness of
the two fluid model in the MHD regime.

%____________________________________________________________________________

\section{Discussion}

In the present paper we have studied a recently presented fluid model
accounting for the electron spin \cite{zamanian2010-2}, and adopted it to
the MHD regime. The main feature of the original model is that in addition
to basic spin effects as the magnetic dipole force, spin precession, and the
magnetization current it incorporates spin-velocity correleations. The
spin-velcoity correlations have been shown to be important for a number of
spin plasma phenomena \cite{zamanian2010-2,stefan2011-1}. Introducing the
approximations appropriate for the MHD regime, it turns out that essentially
the ordinary MHD equations are recovered, but with a magnetization that
needs to be determined. This can be done in different ways. Firstly from a
single fluid model of electrons, that besides the basic spin precession
contains spin-velocity correlations. Or, secondly, from a two-fluid model
where spin-up and spin-down electrons constitute different species. A
theoretical argument is presented in Section II B suggesting that these two
models are equivalent in the MHD regime. However, the equivalence argument
depends on certain assumptions which is difficult to justify rigorously, and
thus practical tests of the equivalence is valuable. For this purpose we
have evaluated the different models using a nonlinear three wave interaction
as a test problem. Specifically, we have computed the coupling coefficients
between two shear Alfv\'en waves and a compressional Alfv\'en wave. A classical
as well as quantum mechanical (spin) contribution to the coupling
coefficients are found, and the coupling coefficients are indeed identical
in the two models. Furthermore, the coupling cofficients obey the
Manley-Rowe symmetries \cite{Manley-Rowe}. The Manley-Rowe relations is a
reflection of the underlying Hamiltonian structure \cite{Larsson99} of the
model. The fact that the coefficients preserve these relations strongly
suggests that the approximations made when deriving the models are sound, as
otherwise it is highly likely that the Manly-Rowe symmetries would be broken.

Since we have put forward two somewhat different models in this paper, one
may ask which one that is most easy to use. For the analytical calculations
made here, the degree of complexity is found to be roughly the same.
However, the two-fluid model has a great advantage in case one would like to
do Particle-In-Cell (PIC) simulations. In that case, the only modification
of a standard code would be to have two species of electrons, and add a
force proportional to $\pm \mu _{e}\nabla B$ in the momentum equation, as
well as to compute $\mathbf{M}=\mu _{e}\mathbf{b}(n_{\uparrow
}-n_{\downarrow })$ to find the contribution from the Magnetization current
in Amperes law. As the variables for the density and magnetic field are
monitored throughout the PIC-simulations anyway, no new equations and only
little extra complexity is added to the general concept. Developing a
PIC-scheme incorporating spin-velocity correlations, however, is a much more
cumbersome project, as the evolution equation for this object is not present
in present PIC-schemes, and also such equations are considerably more
complex. Furthermore, it is not clear that it is at all possible to model
spin-velocity correlations as a single-particle property, which makes the
adaption of this model to the PIC-scheme questionable conceptually. Thus we
conclude that the two-fluid model is valuable for the purpose of
incorporating electron spin effects in PIC-simulations, although we stress
that the applicability of such an approach will be limited to the MHD regime.


\begin{thebibliography}{99}
\bibitem{proton-note} Taking protons as a typical example of ions, we may
note that he proton magnetic moment is of the order of the nuclear magneton,
which is a factor $m_{e}/m_{p}$ smaller than the Bohr magneton.
\bibitem{Magnetization-note} For the purpose of calculating a coupling
coefficient corresponding to a thermodynamic \ equilibrium state, such an
approximation may not be adequate, as we know that there are sightly more
particles in the lower energy spin state, corresponding to a non-zero total
magnetization. However, our chosen background state is nevertheless a valid
dynamic equilibrium, which suffices for the main purpose of this article,
which is to compare the predictions of the two models.
\bibitem{Frequency-note} We have $\omega _{cg}-\omega _{c}=(g/2-1)\omega
_{c}\approx 0.0016\omega _{c}$, which is of the order of the ion-cyclotron
frequency. Thus we have $\omega /(\omega _{cg}-\omega _{c})\ll 1$ within the
standard MHD regime. Accordingly $\omega /(\omega _{cg}-\omega _{c})$ has
been used as an expansion parameter in section II and III, although we in
some cases have kept first order corrections in this parameter for
illustrative purposes.
\end{thebibliography}
\end{document}